\documentclass[twocolumn,10pt,prl,nofootinbib, superscriptaddress]{revtex4}

\usepackage{graphicx}
\usepackage{dcolumn}
\usepackage{bm,amsfonts,amsthm,amsmath,amssymb}
\usepackage[]{epsfig,graphics}
\usepackage{comment}
\usepackage[T1]{fontenc}
\usepackage{lipsum}
\usepackage{ulem}
\usepackage{multirow}
\usepackage{color}
\usepackage{lastpage}
\usepackage{enumerate}
\usepackage{subfigure}
\usepackage{wasysym}
\usepackage{hyperref}
\usepackage{graphicx}

\hypersetup{
    bookmarks=true,         
    unicode=false,          
    pdftoolbar=true,        
    pdfmenubar=true,        
    pdffitwindow=false,     
    pdfstartview={FitH},    
    pdftitle={My title},    
    pdfauthor={Author},     
    pdfsubject={Subject},   
    pdfcreator={Creator},   
    pdfproducer={Producer}, 
    pdfkeywords={keyword1} {key2} {key3}, 
    pdfnewwindow=true,      
    colorlinks=false,       
    linkcolor=red,          
    citecolor=green,        
    filecolor=magenta,      
    urlcolor=cyan           
}

\newenvironment{itemize*}
  {\begin{itemize}
    \setlength{\itemsep}{0pt}
    \setlength{\parskip}{0pt}}
  {\end{itemize}}

\newenvironment{enumerate*}
  {\begin{enumerate}
    \setlength{\itemsep}{0pt}
    \setlength{\parskip}{0pt}}
  {\end{enumerate}}

\newenvironment{description*}
  {\begin{description}
    \setlength{\itemsep}{0pt}
    \setlength{\parskip}{0pt}}
  {\end{description}}

\def\ben{\begin{enumerate*}}
\def\een{\end{enumerate*}}
\def\bi{\begin{itemize*}}
\def\ei{\end{itemize*}}
\def\bd{\begin{description*}}
\def\ed{\end{description*}}
\def\be{\begin{equation}}
\def\ee{\end{equation}}
\def\bea{\begin{eqnarray}}
\def\eea{\end{eqnarray}}
\def\bfl{\begin{flushleft}}
\def\efl{\end{flushleft}}

\newcommand{\gsim}{\lower.7ex\hbox{$\;\stackrel{\textstyle>}{\sim}\;$}}
\newcommand{\lsim}{\lower.7ex\hbox{$\;\stackrel{\textstyle<}{\sim}\;$}}

\newcommand{\beq}{\begin{equation}}
\newcommand{\eeq}{\end{equation}}
\newcommand{\fr}{\frac}

\newcommand{\rhoA}{\rho_A}
\newcommand{\rhor}{\rho_r}

\newcommand{\rhoB}{\rho_B}

\newcommand{\gam}{\Gamma_B}
\newcommand{\dA}{\delta_A}
\newcommand{\dr}{\delta_r}
\newcommand{\dB}{\delta_B}
\newcommand{\tA}{\theta_A}
\newcommand{\tr}{\theta_r}
\newcommand{\tB}{\theta_B}

\newcommand{\rh}{{\mbox{\tiny RH}}}

\begin{document}

\title{ Concentrated Dark Matter: \\
Enhanced Small-scale Structure from Co-Decaying Dark Matter}

\author{Jeff A. Dror}\email{jdror@lbl.gov}
\affiliation{Department of Physics, University of California, Berkeley, CA 94720, USA}
\affiliation{Ernest Orlando Lawrence Berkeley National Laboratory, University of California, Berkeley, CA 94720, USA}
\author{Eric Kuflik}\email{eric.kuflik@mail.huji.ac.il }
\affiliation{Racah Institute of Physics, Hebrew University of Jerusalem, Jerusalem 91904, Israel}
\author{Brandon Melcher}\email{bsmelche@syr.edu}
\author{Scott Watson}\email{gswatson@syr.edu}
\affiliation{Department of Physics, Syracuse University, Syracuse, NY 13244, USA}

\date{\today}

\begin{abstract}
We study the cosmological consequences of co-decaying dark matter -- a recently proposed mechanism for depleting the density of dark matter through the decay of nearly degenerate particles. A generic prediction of this framework is an early dark matter dominated phase in the history of the universe, that results in the enhanced growth of dark matter perturbations on small scales. We compute the duration of the early matter dominated phase and show that the  perturbations are robust against washout from free-streaming. The enhanced small scale structure is expected to survive today in the form of compact micro-halos and can lead to significant boost factors for indirect detection experiments, such as FERMI, where dark matter would appear as point sources.  
\end{abstract}
\maketitle
\thispagestyle{empty}
\section{Introduction}
The thermal history of the universe is established below temperatures around an MeV. Through the precise predictions and measurements employed to study Big Bang Nucleosynthesis (BBN) and the Cosmic Microwave Background (CMB), we can now place stringent bounds on any particles beyond the Standard Model (SM) that were in thermal equilibrium at these times. Conversely, the SM does not provide a 
means of probing its thermal history at temperatures higher than $\mathcal{O}(\rm MeV)$, leading to the common lore that any thermal dynamics above this scale will not be accessible to experiment. Many well-motivated models of dark matter (DM) predict masses above this scale; this makes it challenging to use cosmology to place constraints on the creation mechanism of DM.

Recently, however, it has been proposed that an early period of matter
domination (before BBN) would have observable
implications~\cite{Gelmini:2008sh,Erickcek:2011us,Erickcek:2015jza}.
The idea is that a period of early matter domination would lead to
structure formation in DM prior to freeze-out. These early seeds of
structure can result in {\it concentrated} dark matter, where the bulk
of dark matter is found in  dark compact objects today. This
intriguing possibility has led to a surge of studies of early
structure formation~\cite{Fan:2014zua,Erickcek:2015jza,Choi:2017ncz,Ozsoy:2017hoh,Waldstein:2016blt,Barenboim:2013gya,Giblin:2017wlo,Erickcek:2011us}.
However, while these models do predict significant matter domination
and pre-BBN structure formation, one often finds that the
perturbations do not survive to today. This is due to either DM being
kinetically coupled to the radiation bath during or after the matter
dominated epoch or because reheating washes out the perturbations.

An alternative mechanism which leads to an early matter dominated era and enhanced structure formation, is the recently proposed `co-decaying DM' framework~\cite{Dror:2016rxc}. (For other recent models that include an early period of matter domination in a dark sector see Refs.~\cite{Randall:2015xza,Pappadopulo:2016pkp,Farina:2016llk,Berlin:2016gtr}.) Here, DM itself comes to dominate the total energy density leading to the creation of small-scale structure. Part of the dark sector later decays to SM particles, reheating the SM bath prior to BBN. In this {\it Letter}, we show that since co-decaying DM decouples from all lighter degrees of freedom very early in the history of the universe, the substructure is not washed out by free-streaming or reheating effects. This results in a viable candidate for significantly enhanced small-scale substructure from early universe matter domination.

\section{Co-decaying dark matter}
We begin with a brief review of co-decaying DM, referring the reader to Ref.~\cite{Dror:2016rxc} for more details.
The lightest particles in the dark sector are a (nearly) degenerate species of dark particles, denoted by $A$ and $B$, where $A$ will comprise the DM today, and $B$ is unstable, decaying out of equilibrium. The dark particles thermally decouple from the SM while they are still relativistic in the early universe.  
The two dark sector particles remain in equilibrium with each other via large $AA \leftrightarrow BB$ annihilations, but due to being degenerate and decoupled from the SM, they do not undergo Boltzmann suppression as they become non-relativistic.
Instead, the suppression of the number density occurs when the $B$ particles begin to decay, which results in depletion of the $A$ population. Eventually, the $A $ population drops out of thermal contact with the $B$'s, and $A $ abundance freezes out.

When the dark sector fields become non-relativistic the dark and visible sectors can be described as a system of interacting fluids. The background evolution equations for the energy densities are
\begin{align} 
\label{rhoAB} \rho'_{A}+ \rho'_{B} &= -3  (\rho_{A}+\rho_{B})  - \frac{ \Gamma _B  }{ H} \rho_B , \\
\label{rhoA} \rho'_A &= -3  \rhoA - \frac{\langle \sigma v \rangle}{m H} \left[ \rhoA^2 - \rhoB^2 \right], \\
\label{rhor} \rho'_r &= -4  \rhor + \frac{ \Gamma _B }{ H } \rho _B ,
\end{align}
where $\rho_\alpha$ is the energy density of the respective particles, and $r$ refers to the SM bath. We use primes to denote derivatives with respect to the number of e-folds, $m$ is the mass of $A,B$, and  $\left< \sigma v \right>$ is  the thermally averaged cross-section for $AA \to BB$. For $ s $-wave scattering we can parametrize the thermally averaged cross-section in terms of the zero temperature cross-section, $ \sigma $, 
$
\langle \sigma v \rangle \simeq \sigma \sqrt{({ 16 }/{ \pi }) ({T_{A,B}}/{m})}.
$
The unique dependence on the temperature of the dark sector, $T_{A,B}$, can lead to novel indirect detection signatures~\cite{Dror:2016rxc,Kopp:2016yji,Okawa:2016wrr}.  

In this work, we will assume that dark number changing processes are small and  that `cannibalization' ~\cite{Carlson:1992fn} is absent. In actual models, cannibalization may play a role, even if very minor. We leave the study of the effects of cannibalization on structure growth to future work~\cite{inprogress}. Furthermore, depending on the model realizing the co-decay framework,  kinetic decoupling might occur slightly prior to chemical decoupling, which could effect the computation of the relic density (see Refs.~\cite{DAgnolo:2017dbv,Binder:2017rgn} for related work). However, such model dependencies  are beyond the scope of this work.

\begin{figure}
  \begin{center} 
\includegraphics[trim={0.6cm 0.3cm 0cm  0cm},clip,width=.5\textwidth]{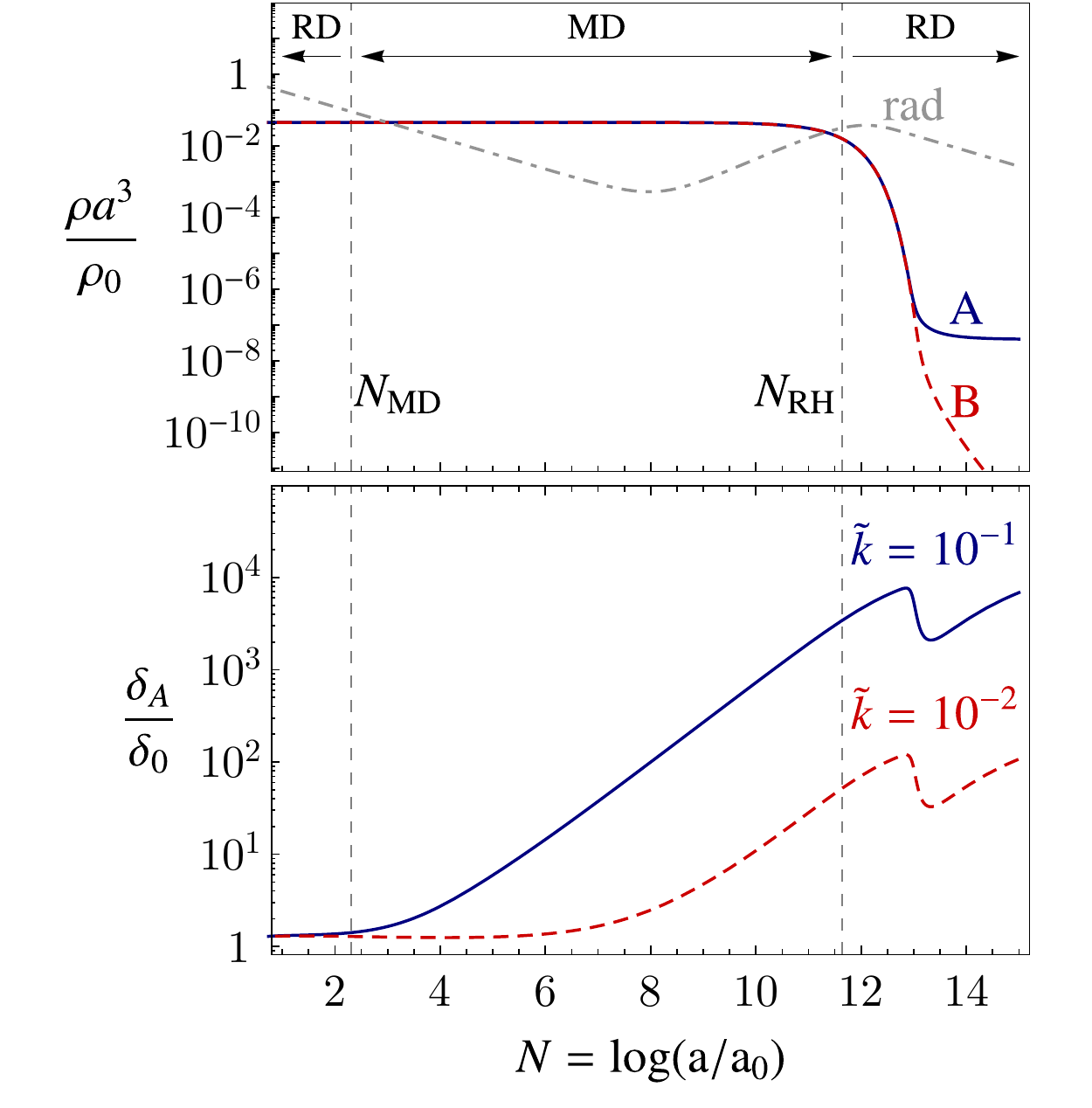} 
\end{center}
\caption{{\bf Top:} Evolution of the background energy densities relative to the total initial density for the benchmark point, $ m = 100~\rm{GeV} $, $ \Gamma _B =   10 ^{ - 22} ~\rm{GeV}   $, and $\xi=0.1$. The subscript 0, refers to values at $ N = 0 $, corresponding to where the dark matter becomes non-relativistic ($ T_{A,B} = m $). Dark matter quickly comes to dominate the energy density of the universe, which lasts until slightly after the onset of $B$ decay. {\bf Bottom:} Evolution of the dark sector density perturbation for mode $\tilde{k}\equiv k/H_{N=0}=10^{-1},\,10^{-2}$, which enter the horizon during matter domination. The modes grows linearly during matter domination, dips during freeze-out, and then grows logarithmically during the radiation dominated epoch.}
\label{fig2}
\end{figure}

The solution to the background energy densities in  Eqs.~(\ref{rhoAB})-(\ref{rhor}) is shown in the top panel of Fig~\ref{fig2}. For a representative benchmark-point with significant matter domination, we take $ m = 100~\rm{GeV} $, $ \Gamma _B =   10 ^{ - 22} ~\rm{GeV}   $, and a ratio of the number of degrees of freedom in the dark sector to the SM at kinetic decoupling  to be $ \xi  =  0.1 $. This scenario corresponds to approximately 9 e-folds of matter domination.

From the time the DM becomes non-relativistic, its density redshifts like matter and quickly comes to dominate the energy density of the universe (marked by $N_{\rm MD}$ in Fig.~\ref{fig2}). Later, $B$ begins to  decay into the SM, reheating the SM bath, and shortly afterwards the universe returns to radiation domination (marked by $N_{\rm RH}$ in Fig.~\ref{fig2}). The period of (dark) matter domination can span many e-folds, during which density perturbations will grow linearly.  The length of matter domination can be split into two parts: the number of e-folds until the onset of $B$ decay (when $ H \sim \Gamma _B $), followed by the time for the decay to deplete the dark sector to the point of a return to radiation domination,
\begin{align} 
N_{\rm RH} - N_{\rm MD}& = \log \frac{ a_{ \Gamma }}{ a_{ \rm MD}} + \log \frac{a _{ \rm RH} }{ a _{ \Gamma }}\,. 
\end{align} 
The subscripts $ \Gamma $ and MD denote values evaluated when $ \Gamma _B  = H $, and at the onset of matter domination, respectively.
The first term can be estimated using entropy conservation prior to decay such that $ a _\Gamma / a _{ \rm MD} = \left( s _{ {\rm SM}} ( a _{ {\rm  MD}} ) / s _{ {\rm SM}} ( a _{ \Gamma } ) \right) ^{ 1/3} $. This can be simplified using the relations of the SM entropy at the onset of matter domination, $n_{A+B} \sim s_{A+B} = \xi  \; s_{\rm SM}$, and at decay, $ \Gamma_B^2 \simeq H^2 \sim m \; n_{A+B} /M_{\rm pl }$. We estimate the second term by tracking the density throughout its decay: $ (H / H _\Gamma ) ^2 = ( a _\Gamma / a ) ^3 e ^{ - \Gamma _B  t /2 } $. Using $ H d t  = d a / a $, this gives a differential equation which can be solved for $ a _{ \rm RH} / a _\Gamma $. In total we find,
\begin{equation} 
N_{\rm RH} - N_{\rm MD} \simeq \frac{1}{3}\log \frac{\xi^4 m^4}{M_{\rm pl }^2 \Gamma_B^2 } + 1  \,.
\end{equation} 

\begin{figure}
  \begin{center} 
\includegraphics[width=.54\textwidth]{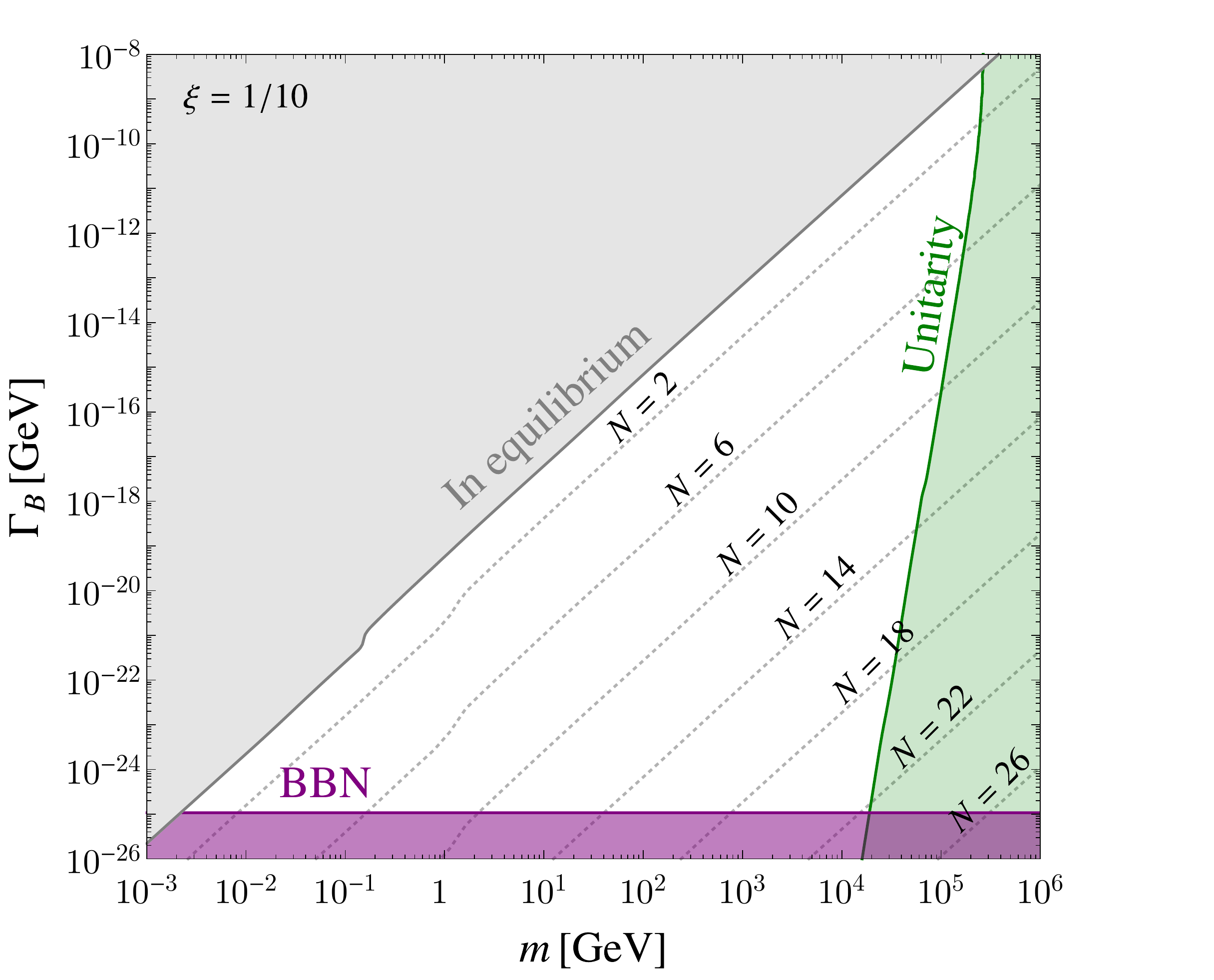} 
\end{center}
\caption{Viable parameter space for co-decaying DM assuming no cannibalization~\cite{Dror:2016rxc}. The different regions show constraints from $\Delta N_{\rm eff}$ ({\bf \color[rgb]{.5,0,.5}purple}); ensuring that DM decays out of equilibrium ({\bf \color[rgb]{0.3,0.3,0.3}gray}); unitarity constraints ({\bf \color[rgb]{0,0.5,0}green}). In addition, there exist model-dependent constraints from indirect detection searches which we omit here. The number of e-folds of matter domination are overlaid, where here $N$ denotes, in shorthand,  $N_{\rm RH} - N_{\rm MD}$.}
\label{fig:MDefolds}
\end{figure}

In Fig.~\ref{fig:MDefolds}, we show the parameter space of co-decaying DM, in terms of $\Gamma_B$ and $m$, and overlay the number of e-folds of matter domination, given that the two sectors were once kinetically coupled early in the universe, and had entropy ratio $ \xi  =  0.1$ at kinetic decoupling. For this choice $\xi$, up to $22$ e-folds of matter domination are possible. Larger values of $\xi$ would increase the duration of the early matter dominated epoch.

\section{Growth of cosmological perturbations and free-streaming}
 We now explore the effects of early matter domination on the growth of co-decaying DM perturbations, where they are expected to grow rapidly. 
In longitudinal gauge, without anisotropic stress, the metric is
\begin{equation} 
ds^2=-\left(1+2 \Phi \right) dt^2 + a(t)^2\left(1-2 \Phi \right) \delta_{ij} dx^i dx^j.
\end{equation} 
Working in momentum space, the time-time component of the perturbed Einstein equation is
\begin{equation} 
\label{EE1}
\left( \frac{k^2}{3 a^2 H^2} + 1 \right) \Phi + \Phi' =-\fr{1}{6H^2m_p^{2}}\sum_{\alpha} \delta \rho_{(\alpha)},
\end{equation} 
where $v_{(\alpha)}, \delta \rho_{(\alpha)}$, $ \delta p_{(\alpha)}$ are scalar velocity, density, and pressure perturbations for each fluid, respectively. 
Introducing the fractional density perturbations $\delta_{(\alpha)} \equiv \delta \rho_{(\alpha)} / \rho_{(\alpha)}$ and defining the velocity perturbation for each fluid as $\theta_{(\alpha)}=a^{-1}\nabla^{2}v_{(\alpha)}$,  the  continuity equations in momentum space are given by
\begin{align} 
\label{dA} \dA' &+ \fr{\tA}{aH} - 3\Phi' = \nonumber \\
&-\fr{\langle \sigma v \rangle}{m H \rhoA} \left[ \rhoA^2 \left( \Phi + \dA \right) - \rhoB^2 \left( \Phi + 2 \dB - \dA \right) \right],  \;\;\;\;\;\; \\
\label{dB} \dB' &+ \fr{\tB}{aH} - 3\Phi' = - \fr{\gam}{H} \Phi  \nonumber \\
&+ \fr{\langle \sigma v \rangle}{m H \rhoB} \left[ \rhoA^2 \left( \Phi + 2 \dA - \dB \right) - \rhoB^2 \left( \Phi + \dB \right) \right], \;\;\;\;\;\; \\
\label{dr} \dr' &+ \frac{4}{3} \fr{\tr}{aH} - 4\Phi' =  \frac{\gam}{H} \frac{\rhoB}{\rhor} \left[ \Phi + \dB - \dr \right]\,.
\end{align} 
Similarly, the equations for the velocity perturbations are
\begin{align}  
\label{tA} \tA' + \tA - \fr{k^2}{aH} \Phi &= \fr{\langle \sigma v \rangle}{m H \rhoA} \left[ \rhoB^2 \left( \tB - \tA \right) \right], \\
\label{tB} \tB' +\tB - \fr{k^2}{aH} \Phi &= \fr{\langle \sigma v \rangle}{m H \rhoB} \left[ \rhoA^2 \left( \tA - \tB \right) \right], \\
\label{tr} \tr' - \fr{k^2}{aH} \left( \fr{\dr}{4} +\Phi\right) &=  \frac{\gam}{H} \frac{\rhoB}{\rhor} \left[ \frac{3}{4} \tB - \tr \right].
\end{align}
We took each fluid to have a definite equation of state with pressure $p_{(\alpha)}=w_{(\alpha)}\rho_{(\alpha)}$ and hence $\delta p_{(\alpha)}= c_{s(\alpha)}^{2} \delta\rho_{(\alpha)}$ with $c_{s(A)}^2 = c_{s(B)}^2 = 0, c_{s(r)}^2 =1/3$. 
This set of differential equations can be closed using Eq.~\eqref{EE1}.\footnote{In principle the inevitable dark temperature dependence of the thermally averaged cross-section requires the inclusion of an additional first order perturbation equation. Since the dark temperature around freeze-out may not be well understood in some models realizing the co-decay framework we do not include these effects here, but emphasize the schematic nature of the numerical solutions. Furthermore, the annihilations terms only slightly effect the perturbations, thus the  inclusion dark temperature perturbations will have a small effect on the final results. We have checked this explicitly.} We take adiabatic initial conditions\footnote{For a situation with non-adiabatic initial conditions in a matter phase we refer to \cite{Iliesiu:2013rqa}.} for the perturbations as in \cite{Erickcek:2011us,Fan:2014zua}.

Our solutions for the linear density perturbations are presented in the  bottom panel of Fig.~\ref{fig2}. Here we consider modes, $k=0.1 \times H_{N=0}$ and $k=0.01 \times H_{N=0}$, which enter the horizon at $N \simeq 2.7$ and  $N \simeq 6.7$, respectively. The over-density grows linearly during matter domination, until reheating ($N_{RH }\simeq 13$), at which point the perturbations are slightly washed out by $ A A \rightleftharpoons BB $ annihilations during freeze-out, resulting in the dip in Fig.~\ref{fig2}. 
  After that, the perturbations grow logarithmically during the radiation dominated era.   

The amplitude of general DM perturbations can be estimated analytically by approximating the decay as instantaneous and the dark matter as a single fluid. Perturbations that enter the Hubble radius prior to reheating ($k>k_{\rh}$), take the form 
\begin{equation}  \label{thisresult}
\left| \delta_A \right|  = \frac{2}{3} \left( \frac{k}{k_\rh} \right)^2 \Phi_0 \left[ 1 + \ln\left( \frac{a}{a_\rh} \right) \right],
\end{equation} 
where $k_\rh$ is given by
\begin{equation}   \label{kreheat}
k_\rh \equiv a_\rh H_\rh= 0.1~\mbox{pc}^{-1} \times \left( \frac{T_\rh}{3 \; \mbox{MeV}} \right) \left( \frac{g_\ast(T_\rh)}{10.75} \right)^{1/6},
\end{equation} 
with $T_\rh$ the temperature of the SM bath immediately proceeding decay. Eq.~\eqref{thisresult} is found by accounting for the initial amplitude of the perturbations as a mode 
enters the horizon, which grows linearly during matter domination, and then logarithmically with the scale factor after reheating. We note that the above expressions agree with those of Ref.~\cite{Erickcek:2015jza}.

The scale of reheating, given by Eq.~\eqref{kreheat}, should be compared to the scales of kinetic decoupling and free-streaming, since both can wash out structure. Early kinetic decoupling is a defining feature of the framework and therefore there is no collisional damping to suppress growth of structure in co-decaying DM. 
On the other hand, the free-streaming wavelength is given by 
\begin{equation} 
 \frac{1}{k_{\rm FS}} = \int_{a_{kd}}^{a_{\rm eq}} \frac{\langle v \rangle}{H a^2} da.
\end{equation} 
where $\langle v \rangle$ is the average velocity of the dark  $A+B$ fluid.
Immediately after kinetic decoupling, the free-streaming dark matter fluid is relativistic, $\langle v \rangle  = 1$. Then, close to when  the universe becomes  matter dominated, the DM becomes non-relativistic  and its velocity begins to redshift like the scale factor, and quickly slows. Thus most of the free-streaming takes place before the matter dominated phase begins. Therefore, for co-decaying DM, the free-streaming length is roughly the size of the horizon at the time of matter domination, $k_{\rm FS} \simeq k_{\rm MD}$. Free-streaming will only potentially wash out modes that have just entered the horizon after matter domination, and most modes will grow unimpeded by free-streaming effects. 

Relative to the scale at reheating,
\begin{equation} 
\frac{k_{\rm FS}}{k_{\rm RH}} \simeq\frac{k_{\rm MD}}{k_{\rm RH}} \simeq \left(\frac{g_{*,\rm MD}}{g_{*,\rm RH}}\right)^{1/6} \left(\frac{m{\xi}}{T_{\rm RH}}\right)^{2/3}. \label{kmd}
\end{equation}
 The free-streaming damping scale ($k_{\rm FS} \simeq k_{\rm MD}$)  
 will set the critical scale $k_{\rm cut} =  k_{\rm MD}$ for the smallest size of sub-halos. In Ref.~\cite{Erickcek:2011us}, it was found that whichever scale sets the cutoff $k_{\rm cut}$ (in general $k_{\rm FS}$ or $k_{\rm KD}$, whichever is smaller) needs only satisfy $k_{\rm cut} / k_\rh \gtrsim 10$ in order for the early matter epoch to lead to enhanced small scale structures (micro-halos). In co-decaying DM, this is typically orders of magnitude larger ({\it e.g.}, $\mathcal{O}(10^3)$ for the parameters in Fig.~\ref{fig2}).

The survival of small-scale structure in co-decaying DM is in contrast to the case of dark matter produced from the decays of an out of equilibrium particle, such as moduli decaying to a neutralino. Indeed, in Ref.~\cite{Fan:2014zua} it was shown in this context that the enhanced growth of structure would not survive either due to late time kinetic decoupling, or  free-streaming of relativistic DM produced from moduli decays.

\section{Present day substructure}
Having established that the enhanced perturbations for the growth of micro-halos can survive reheating and wash-out effects, we now consider the possible implications for structure formation today. Following  Refs.~\cite{Erickcek:2011us,Erickcek:2015jza}, who use the Press-Schechter formalism~\cite{Press:1973iz}, we now comment on the qualitative features of the predicted substructure. 
 
Refs.~\cite{Erickcek:2011us,Erickcek:2015jza}  established the fraction of  the dark matter abundance found in micro-halos today resulting  from an early matter dominated phase. There, it was found that the effect of the enhanced growth can be captured by altering the transfer function on the relevant scales, namely between the beginning of matter domination and the time of reheating. Additionally, the growth function will differ from the standard scenario since baryons will not play a role in structure formation at these scales.  

From the Press-Schechter formalism we expect that once the rms density perturbation exceeds the critical value $\delta_c =1.69$, compact micro-halos will form. 	Refs.~\cite{Erickcek:2011us,Erickcek:2015jza} found that the  rms density perturbation is rather insensitive to the reheat temperature, but depends critically on $k_{\rm cut} / k_\rh$.  If this ratio exceeds $\mathcal{O}(10)$ then micro-halos can form; importantly, this ratio is orders of magnitudes larger in co-decaying DM. 
In particular, it was shown that this ratio not only determines the masses of the micro-halos but 
also their time of formation.  As the $k_{\rm cut} / k_\rh$ ratio increases, the redshift at which these micro-halos form increases as well.
Moreover, the higher the value of the ratio, the more peaked the mass distribution of the micro-halos is towards the largest possible size. 
The largest micro-halos are set by the size of the horizon at reheating, $1/k_{\rm RH}$, and will have masses smaller than or near
\begin{equation} 
  \label{massrh}
M_\rh \equiv \frac{4}{3} \pi \rho_{A}^{(0)} k_\rh^{-3}  
\simeq  10^3 M_{\oplus}\; \left( \frac{3 \; \mbox{MeV}}{T_\rh} \right)^3 \left( \frac{10.75}{g_\ast(T_\rh)} \right)^{1/2}.
\end{equation} 
Namely, for reheating near BBN, co-decaying DM can produce micro-halos with masses as large as a thousand Earth masses or less. The combination of the above with the natural prediction in the co-decaying DM framework of  $k_{\rm cut}  / k_\rh \gg 1$,  
suggests that co-decaying DM will lead to the formation of micro-halos with masses peaking around the value given in Eq.~\eqref{massrh}.

If these structures survive until today, they lead to high concentrations of DM which result in large boost factors for the self-annihilation of DM~\cite{Erickcek:2015jza}.  Thus, co-decaying DM predicts enhanced signals in indirect-detection experiments.
Furthermore, sub-halos of the size predicted here would appear as point sources, and could give a DM explanation~\cite{Agrawal:2017pnb} to the unidentified point sources observed by FERMI-LAT~\cite{Lee:2015fea,Bartels:2015aea,Fermi-LAT:2017yoi}. 

We conclude with several comments and open questions. First, N-body simulations are most likely needed in order to evaluate the survival rates of the micro-halos from high red-shifts until today. Next, it would be interesting to study the internal structure of these halos;  given the high value of $k_{\rm cut}$ in co-decaying DM, the distribution may be fairly homogeneous -- i.e. the dark matter would be quite concentrated. Finally, the early matter dominated phase can also result in the formation of primordial black holes~\cite{Georg:2017mqk,Georg:2016yxa}, which can provide another component of the cosmological DM, in addition to the co-decaying DM. We leave the exploration of these important questions to future work. 
 
\section*{Acknowledgements}
We thank Avishai Dekel, Cosmin Isle, Adrienne Erickcek, Jonah Kudler-Flam,  Michael Geller, and Gustavo Marques Tavares for useful discussions, and Yonit Hochberg for comments on the manuscript. SW and BM were supported in part by NASA Astrophysics Theory Grant NNH12ZDA001N and DOE grant DE-FG02-85ER40237. EK is supported by the I-CORE Program of the Planning Budgeting Committee (grant No. 1937/12), the Israel Science Foundation (grant No. 1111/17), and the Binational Science Foundation (grant No. 2016153).  EK and SW thank hospitality of the Aspen Center for Physics, which is supported by NSF grant PHY-1066293. JD is supported in part by the DOE under contract DE-AC02-05CH11231.

%

\end{document}